\documentclass[twoside,aps,10pt,twocolumn]{revtex4-1}
 
\usepackage{amsmath}
\usepackage{amssymb}
\usepackage{amsfonts}
\usepackage{graphicx}
\usepackage{soul}
\usepackage{braket}

\newcommand{\ee}{\mathrm{e}}
\newcommand{\ii}{\mathrm{i}}

\newcommand{\bvec}[1]{\mathbf{#1}}

\begin{document}
	
	\title{Angular-Momentum Couplings in Ultra-Long-Range Giant Dipole Molecules}

	\date{\today}
	
	\author{Thomas Stielow}
	\author{Stefan Scheel}
	\author{Markus Kurz}
	\affiliation{Institut f\"ur Physik, Universit\"at Rostock, Albert-Einstein-Stra{\ss}e 23, D-18059 Rostock, Germany}

	\begin{abstract}
		In this article we extend the theory of ultra-long-range giant dipole molecules, formed by an atom in a giant dipole state and a ground-state alkali atom, by angular-momentum couplings known from recent works on Rydberg molecules. In addition to $s$-wave scattering, the next higher order of $p$-wave scattering in the Fermi-pseudopotential describing the binding mechanism is considered. Furthermore, the singlet and triplet channels of the scattering interaction as well as angular-momentum couplings such as hyperfine interaction and Zeeman interactions are included. Within the framework of Born–Oppenheimer theory, potential energy surfaces are calculated in both first-order perturbation theory and exact diagonalization. Besides the known pure triplet states, mixed-spin character states are obtained, opening up a whole new landscape of molecular potentials. We determine exact binding energies and wave functions of the nuclear rotational and vibrational motion numerically from the various potential energy surfaces.
	\end{abstract}

	\maketitle

	\section{Introduction}
	Rydberg atoms are a topic of special interest not only in atomic physics but also in in quantum optics, where they serve as a sensitive tool for electromagnetic detection down to the single-photon level \cite{Haroche_1996,Haroche_2007}. Rydberg states with orbit sizes of several $\mu\text{m}$ \cite{Pfau_n200} are some of the largest quantum systems known in the framework of ultracold atomic gases. The huge orbit sizes, often larger than the free path length of the surrounding cloud of atoms inside gas cells, open up possibilities for secondary atoms to directly interact with the loosely bound Rydberg electron. The effects of such interactions have been analyzed in experiments back in 1933 independently by Amaldi and Segr\'{e} in Rome \cite{Amaldi_1934} and by the group of F\"uchtbauer in Rostock \cite{Fuechtbauer_1933,Fuechtbauer_1934,Fuechtbauer_1934_Kalium}. They observed line shifts towards the red and, surprisingly, also towards the blue. This led to the development of the modern quantum scattering theory by Fermi \cite{Fermi_1934}, who assumed the electron as a quasi-free particle with low momentum that scatters on a secondary perturber atom.
	
	More than half a century later, the question whether the scattering interaction may lead to a binding interaction was addressed by Greene \textit{et al.} \cite{Greene_Trilobite}. Previously, it was shown that the scattering interaction in partial-wave expansion can be modeled by a delta-shaped contact potential \cite{Omont_1977}. By considering only the leading order of $s$-wave scattering they predicted two different types of bound molecular states between a rubidium Rydberg and ground state atom. The first species originating from low angular momentum states was found experimentally in 2009 \cite{Pfau_2009low}. The second type are the so-called trilobite states, named after the shape of their electronic probability density. They were detected in 2015 in cesium \cite{Booth_2015} and in 2017 in rubidium ultracold atomic gases \cite{Kleinbach_2017_trilobite}. In rubidium, the next higher order in the partial-wave expansion of the interaction potential, the $p$-wave scattering, shows a shape resonance at a certain electron momentum \cite{Fabrikant_2002}. Consequently, it may dominate over the leading scattering order, resulting in a new type of Rydberg molecules known as butterfly states \cite{Greene_Butterfly,Niederpruem_2016_butter}. 

	A further variable in the scattering interaction is the combined spin of scattered electron and perturber shell. Earlier works on Rydberg molecules only considered a triplet configuration which is easy to preparate in the laboratory, as the single $s$-wave scattering is repulsive for low momenta. Including these angular momentum couplings \cite{Anderson_2014} led to a wide range of discoveries including new excitation channels \cite{Kleinbach_2017_trilobite}, indirect measurements of the singlet scattering length \cite{Boettcher_2016}, and observations of spin flips in Rydberg molecules \cite{Niederpruem_2016_spin}.
	
	An exotic species of Rydberg atoms, that can exist only in the presence of external crossed electric and magnetic fields, are the so-called giant dipole (GD) atoms. First introduced theoretically in the 1970s for hydrogen-like systems such as excitons \cite{Burkova_1976}, major steps in their understanding were made in the early 1990s \cite{Baye_1992,Dzyaloshinskii_1992,Schmelcher_1993}. A full theory was developed by Dippel \textit{et al.} \cite{Dippel_Giant_Dipole}, and experimental observations have been claimed in the following years \cite{Raithel_1993,Fauth_1987}. In contrast to usual Rydberg states, the electronic probability density is strongly decentered and directed, resulting in a huge permanent electric dipole moment. Dippel \textit{et al.} found that the underlying Hamiltonian may be transformed into a gauge-invariant one-particle problem leading to an effective potential possessing an outer well at distances of several hundred thousand Bohr radii relative to the Coulomb center \cite{Dippel_Giant_Dipole}. This outer well supports loosely bound states in which the electron, similar to a usual Rydberg state, possesses only low kinetic energy. Due to the analogy to Rydberg atoms, the possibility of molecules formed by GD states with a neutral ground-state perturber via scattering interaction was studied by Kurz \textit{et al.} \cite{Markus_erstes}. Similar to the first approach of Greene \textit{et al.} to Rydberg atoms \cite{Greene_Trilobite}, they only considered the triplet $s$-wave scattering. By these means, molecular potential energy surfaces were calculated, predicting the existence of ultra-long-range giant-dipole molecules.
	
	In this article, we extend the theory of ultra-long-range giant-dipole molecules to include higher orders of the interaction potential as well as different spin configurations. We start in Sec.~\ref{sec:molHam}with a molecular Hamiltonian that includes both $s$- and $p$-wave interaction potentials in triplet and singlet configuration as well as all relevant spin and angular momentum couplings of the molecular system. In Sec.~\ref{sec:spinSpace}, we analyze the spin subspace of the molecular system in the regime of different field strengths. We introduce a cylindrical approximation to the GD system in Sec.~\ref{sec:Meth} and illustrate the calculation of molecular states in Born--Oppenheimer approximation. Characteristic potential energy surfaces are presented in Sec.~\ref{sec:PES}, and in Sec.~\ref{sec:Rovi} we calculate the corresponding vibrational states to obtain molecular binding energies. A summary is given in Sec.~\ref{sec:conc}. Atomic units will be used throughout this paper if not stated otherwise.
	
	\section{Molecular Hamiltonian and Interactions}\label{sec:molHam}
	We consider a highly excited hydrogen atom interacting with a neutral ground-state perturber atom (in our case $^{87}$Rb) in crossed static homogeneous electric and magnetic fields $\bvec E$ and $\bvec B$, respectively. The hydrogen atom is assumed to be in a GD state. The full system is described by the Hamiltonian
	\begin{equation}
		H = \frac{\bvec{p}^{2}_\text{n}}{2m_\text{n}} + H_\text{GD} + H_\text{spin} + V_\text{GD,n} \, \text ,\label{eq:molHam}
	\end{equation}
	where the first term is the kinetic energy of the neutral perturber (subscript ``n''), followed by the giant-dipole Hamiltonian of the hydrogen atom interacting with the external fields. The last two terms are interaction terms, where the first includes the intrinsic spin and angular momentum couplings inside the constituents as well as to the external fields, and the last term contains the interaction between perturber and hydrogen atom.
	
	The hydrogen atom in crossed external electric and magnetic fields has been discussed in detail in Ref.~\cite{Dippel_Giant_Dipole}. There, it was shown that the giant-dipole Hamiltonian can be transformed into an effective single-particle problem. The particle in question interacts with the magnetic field and a generalized potential $V$ that depends parametrically on the external fields and the pseudomomentum $\bvec{K}$ and contains both the motional and external Stark terms
	\begin{equation}
		H_\text{GD} = \frac{1}{2 \mu} \left( \bvec{p} - \frac{q}{2} \bvec{B} \times \bvec{r} \right)^2 + V(\bvec{r}) 
	\end{equation}
	with $\mu = m_e m_p / (m_e + m_p)$ and $q = (m_e - m_p) / (m_e + m_p)$. The quantities $e, m_e, m_p$ are the electron charge and mass and the proton mass, respectively, $\bvec{r}$ and $\bvec{p}$ represent the coordinate and canonical momentum of the Rydberg electron. For sufficiently strong electric and magnetic fields, the giant-dipole potential exhibits an outer well with a local minimum at  $\bvec r_0 = (x_0, 0, 0)$ with $x_0\approx-\frac K B+\frac{MK}{K^3-2MB}$ \cite{Dippel_Giant_Dipole}, which is found to be an appropriate approximation to the exact solution given in Ref.~\cite{Schmelcher_Positronium}. Throughout this article, we will consider field strengths of $B=2.35\,\text{T}$ and $E=2.8\times 10^3$V/m (in atomic units: $B=10^{-5}$, $K=1.0$), resulting in a separation of $|x_0| \approx 10^{5} \, \text a_0$. This separation gives rise to a huge electric dipole moment of many tens of thousand Debye. 
	
	In the vicinity of $\bvec r_0$, the potential $V(\bvec r)$ is nearly parabolic and may be approximated in a power series up to $\bvec r^2$. By replacing $\bvec r \to \bvec r + \bvec r_0$, one obtains the potential of a three-dimensional harmonic oscillator \cite{Schmelcher_1993,Dippel_Giant_Dipole}
	\begin{equation}
		V_\text{GD}(\bvec r) = \frac \mu 2 \omega_x^2 x^2 + \frac \mu 2 \omega_y^2 y^2 + \frac \mu 2 \omega_z^2 z^2 \, \text , \label{eq:Vgd}
	\end{equation}
	with eigenfrequencies $\omega_{x,y}^2 = \frac 2 \mu \left(\frac{B^2}{2M} \pm \frac{1}{x_0^3}\right)$  and $\omega_z^2 = -\frac{1}{\mu x_0^3}$.

\begin{figure}[bt]
		\includegraphics[width=\columnwidth]{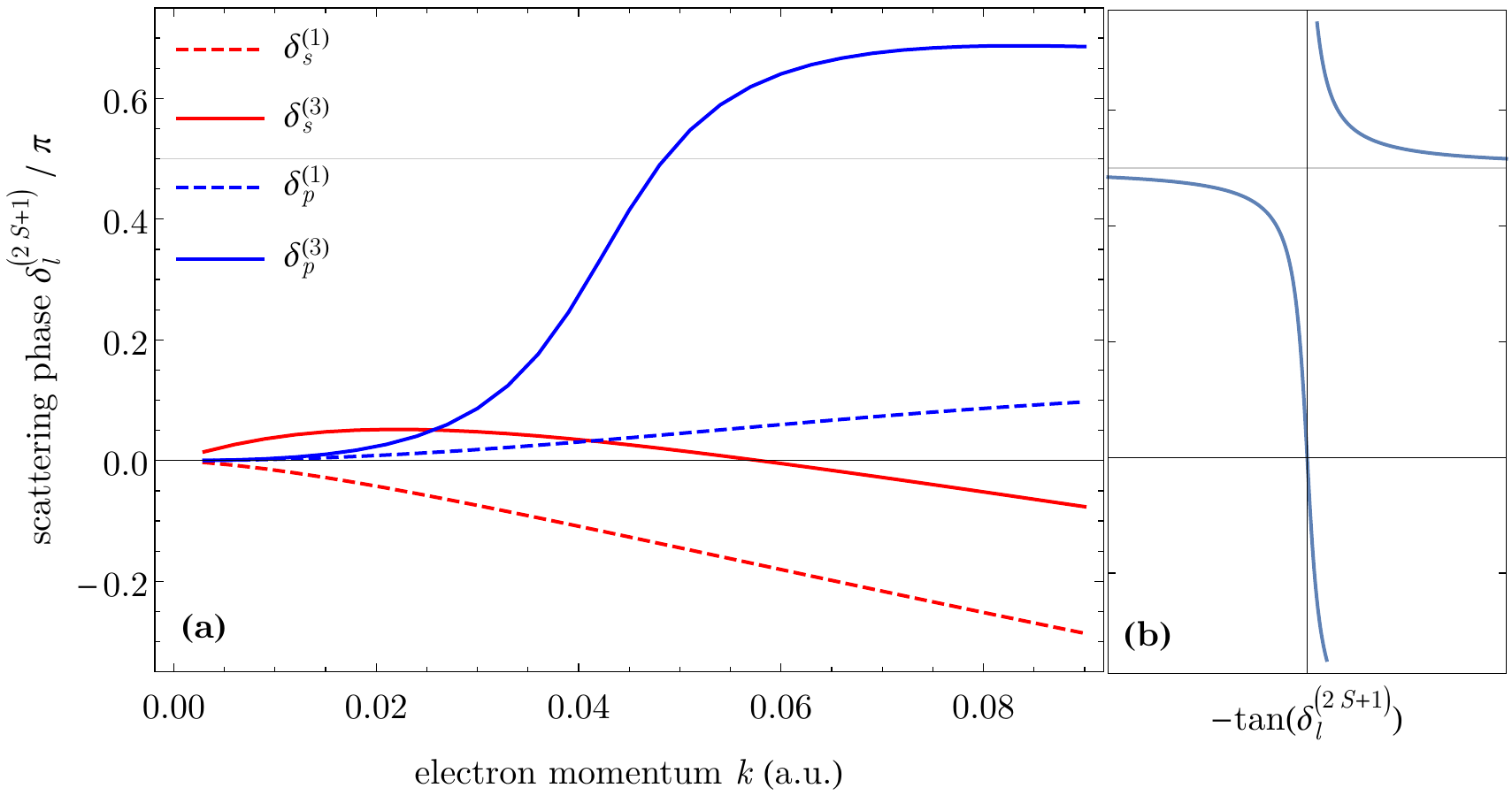}
		\caption{Scattering phases $\delta_l^{(2S+1)}$ of a quasi-free electron on a rubidium ground-state atom (a). The effective scattering length $-\tan[\delta_l^{(2S+1)}]$ (b) diverges at $\delta = \pi / 2$, resulting in a shape resonance observed for $\delta_p^{(3)}$.}\label{fig:scatteringPhases}
	\end{figure}

	In Ref.~\cite{Markus_erstes} it was shown that the interaction between a GD atom and a neutral ground-state perturber atom may be described by a contact potential in the form of an $s$-wave pseudopotential \cite{Omont_1977,Greene_Trilobite}. Recent studies on Rydberg atoms utilizing the same interaction potential have revealed shape resonances in $p$-wave couplings \cite{Greene_Butterfly,Fabrikant_2002,Niederpruem_2016_butter} and dependencies on the spin configuration of electron and perturber \cite{Anderson_2014, Boettcher_2016, Niederpruem_2016_spin}. The corresponding interaction potential then reads as \cite{Anderson_2014}
	\begin{align}
		V_\text{GD, n}(\bvec{r}, \bvec{R}) = & \sum_{S = 0, 1}\left[ 2 \pi A^{(2S + 1)}_{s} \delta(\bvec{r} - \bvec{R}) \right.\nonumber \\
		&\left. + 6 \pi A^{(2S+1)}_{p} \bvec \nabla \delta(\bvec{r} - \bvec{R}) \bvec \nabla \right] \mathbb P_{2S+1} \, \text , \label{eq:Vint}
	\end{align}
	where $\bvec r$ is the electron and $\bvec R$ the perturber position, respectively. In this partial wave expansion the interaction amplitudes are expressed in terms of the scattering lengths $A_{l}^{(2S+1)}$ with $l=S,P$, which are related to the scattering phases by $A_{l}^{(2S+1)}=-\tan[\delta_{l}^{(2S+1)}(k)]/k^{2l+1}$ \cite{Anderson_2014}. For a rubidium perturber, these scattering phases have been calculated in Ref.~\cite{Fabrikant_2002} and are illustrated in Fig.~\ref{fig:scatteringPhases}. 

	In $s$-wave approximation, the triplet spin configuration results in an attractive scattering interaction for low momenta whereas the singlet interaction is repulsive \cite{Greene_Trilobite}. The importance of including the next higher order of scattering stems from the fact that the $p$-wave scattering phase in triplet configuration crosses the value of $\pi/2$ at around $k = 0.05 \, \text{a.u.}$, resulting in a shape resonance \cite{Greene_Trilobite} shown in Fig.~\ref{fig:scatteringPhases}(b). The projection of the corresponding spin singlet ($S=0$) and triplet ($S=1$) configurations of the electron spins $\bvec s_1$, $\bvec s_2$ of the GD atom and the perturber, respectively, is performed by projection operators $\mathbb P_{1,3}$ with $\mathbb P_3=\hat{\bvec s}_1 \cdot\hat{\bvec s}_2+3/4$ and $\mathbb P_1=\mathbb I-\mathbb P_\text{T}$ \cite{Anderson_2014}. 
	
	The residual spin and angular momentum couplings are summarized in the spin Hamiltonian
	\begin{equation}
		H_\text{spin} = A \, \bvec{I}_2 \cdot \bvec{s}_2 + \frac{g_s}{2} B(m_{s_1} + m_{s_2}) + \frac{g_I}{2}BI_{z} \, \text . \label{eq:Hspin}
	\end{equation}
	Here, the first term is the hyperfine interaction of the ground-state perturber atom, with coupling constant $A=3.417\,h\,\text{GHz}$ for ${}^{87}\text{Rb}$ \cite{rb87data}. We neglect the hyperfine coupling of the GD atom due to the huge spatial electron-core separation. The second term in Eq.~\eqref{eq:Hspin} is the normal Zeeman interaction of both the rubidium and hydrogen valence electrons with Land\'{e} factor $g_s \approx 2$. The coupling between the angular momentum of the hydrogen atom's electron is already included in $H_\text{GD}$ while the rubidium ground state $5^2S_{1/2}$ has zero angular momentum. The last term is the Zeeman interaction of the nuclear spin of the perturber in the external magnetic field.

	\section{Spin Space Analysis}\label{sec:spinSpace}
	In spin space, the GD-perturber system is composed of the electronic spins of GD and perturber atom $\bvec s_1$ and $\bvec s_2$, respectively, and the perturber nuclear spin $\bvec I_2$ with $I_2 = 3/2$. While the spin Hamiltonian in Eq.~\eqref{eq:Hspin} conserves $\bvec F_2=\bvec I_2+\bvec s_2$ and $\bvec s_1$, the interaction potential \eqref{eq:Vint} conserves only the total spin $\bvec S=\bvec s_1+\bvec s_2$. Hence, the total Hamiltonian \eqref{eq:molHam} does neither conserve $\bvec F_2$ nor $\bvec S$. Instead, only the sum of all magnetic quantum numbers $\mathcal M = m_{I_2} + m_{s_1} + m_{s_2}$ is a good quantum number. 
	
	Table~\ref{tab:fullM} lists all possible spin states according to the values of $\mathcal M$ using the notation $\left |m_{I_2} \middle |m_{s_1} m_{s_2}  \right>$ with the symbol $\uparrow$ denoting $m_i=+1/2$ and $\downarrow$ representing $m_i=-1/2$.
	\begin{table}[htb]%
		\centering%
		\begin{tabular}{ccccccc}\hline\hline%
			$\mathcal M$&	$+\frac 5 2$&	$+\frac 3 2$&	$+\frac 1 2$&	$-\frac 1 2$&	$-\frac 3 2$&	$-\frac 5 2$\\\hline%
			&$\left |+\frac 3 2 \middle | \uparrow \uparrow \right>$&
			$\left |+\frac 1 2 \middle | \uparrow \uparrow \right>$&
			$\left |-\frac 1 2 \middle | \uparrow \uparrow \right>$&
			$\left |-\frac 3 2 \middle | \uparrow \uparrow \right>$&&\\%	

			&&$\left |+\frac 3 2 \middle | \downarrow \uparrow \right>$&
			$\left |+\frac 1 2 \middle | \downarrow \uparrow \right>$&
			$\left |-\frac 1 2 \middle | \downarrow \uparrow \right>$&
			$\left |-\frac 3 2 \middle | \downarrow \uparrow \right>$\\%

			&&$\left |+\frac 3 2 \middle | \uparrow \downarrow \right>$&
			$\left |+\frac 1 2 \middle | \uparrow \downarrow \right>$&
			$\left |-\frac 1 2 \middle | \uparrow \downarrow \right>$&
			$\left |-\frac 3 2 \middle | \uparrow \downarrow \right>$\\%

			&&&$\left |+\frac 3 2 \middle | \downarrow \downarrow \right>$&
			$\left |+\frac 1 2 \middle | \downarrow \downarrow \right>$&
			$\left |-\frac 1 2 \middle | \downarrow \downarrow \right>$&
			$\left |-\frac 5 2 \middle | \downarrow \downarrow \right>$\\\hline\hline%
		\end{tabular}%
		\caption{The spin basis of the GD-perturber system in the notation of $\left |m_{I_2} \middle | m_{s_1} m_{s_2} \right>$ can be arranged according to the good quantum number $\mathcal M = m_{I_2} + m_{s_1} + m_{s_2}$. The outermost states are pure triplet spin states.}\label{tab:fullM}%
	\end{table}%
	States with $\mathcal M=\pm 5/2$ are pure triplet states, this configuration was already analyzed in Ref.~\cite{Markus_erstes}. The cases with $\mathcal M=\pm 3/2$ can be seen as being already included in the configurations with $\mathcal M=\pm 1/2$. Hence, it is enough to consider only $\mathcal M=+1/2$. 
	
	Expressed in a basis of spin states in the notation used in Table~\ref{tab:fullM}, the spin Hamiltonian \eqref{eq:Hspin} takes the form of a $(2\times 2)\otimes(2\times 2)$ block-diagonal matrix
	\begin{widetext}
	\begin{equation}
		\mathbf{H}_\text{spin} = 
		\begin{pmatrix}
			-\frac{3}{4}A - \frac{g_2}{2} B + \frac{3 g_I}{4} B & \frac{\sqrt{3}}{2}A & 0 & 0 \\
			\frac{\sqrt{3}}{2}A & \frac{1}{4}A+\frac{g_I}{4}B & 0 & 0 \\
			0 & 0 & -\frac{1}{4}A + \frac{g_I}{4}B & A \\
			0 & 0 & A & -\frac{A}{4} + \frac{g_2}{2} B -\frac{g_I}{4}B \\
		\end{pmatrix}\, \text ,	 \label{eq:HspinMat}
	\end{equation}
	\end{widetext}
	which can be diagonalized analytically. The two blocks describe the spin flip of $\bvec s_2$ caused by the hyperfine interaction while the GD spin $\bvec s_1$ is fixed. Consequently, each eigenstate of $H_\text{spin}$ may only be constructed from basis states $\left|m_{I_2}\middle |m_{s_1} m_{s_2}\right>$ sharing the same GD spin $\bvec s_1$.

	\begin{figure}[thb]
		\includegraphics[width=\columnwidth]{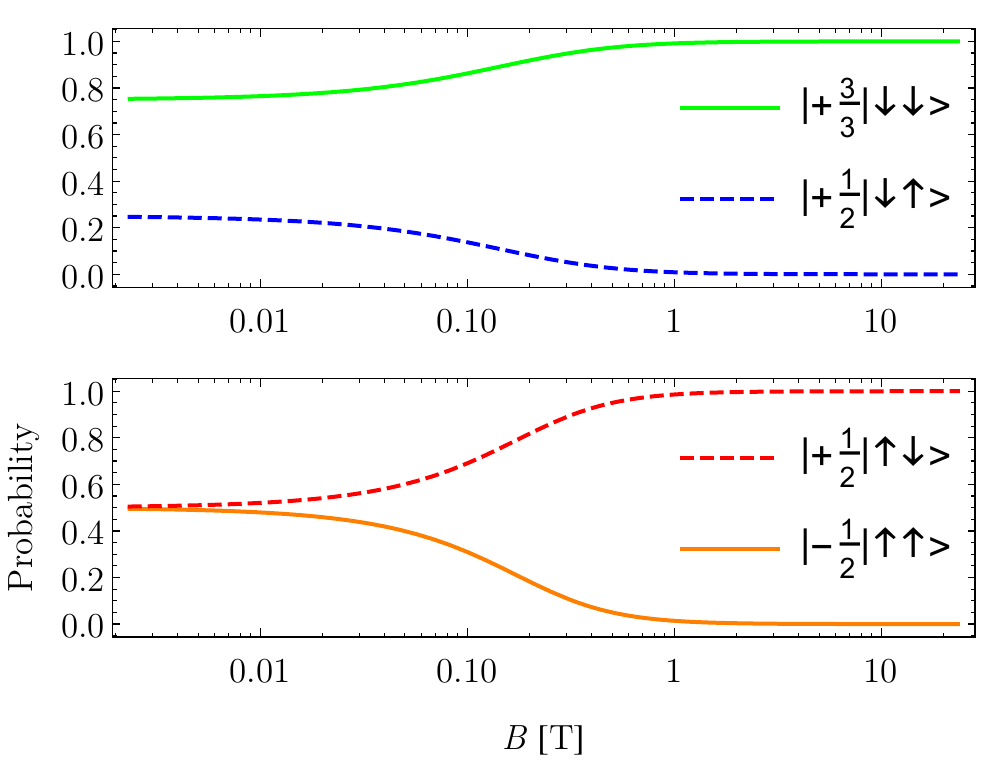}
		\caption{The eigenstates of $H_\text{spin}$ are superpositions of two basis states $\left |m_{I_2} \middle | m_{s_1} m_{s_2} \right>$ with the same $m_{s_1}$. We show the varying state content as a function of the magnetic laboratory field strengths. In weak magnetic fields, the hyperfine interaction in Eq.~\eqref{eq:Hspin} dominates, with the eigenstates corresponding to those of $\bvec I_2 \cdot \bvec s_2$. In strong magnetic fields, the electronic Zeeman term dominates, resulting in eigenstates identified by the total spin magnetic quantum number $M=m_{s_1}+m_{s_2}$. }\label{fig:spinState}
	\end{figure}

	In Fig.~\ref{fig:spinState}, the decomposition of one eigenstate each for both orientations of $\bvec s_1$  is shown over a range of relevant laboratory field strengths. The two remaining eigenstates are fomed by the opposite assignment of the involved basis states. In the regime of weak magnetic field $\lesssim 10\,\text{mT}$, $H_\text{spin}$ is dominated by the hyperfine interaction and the eigenstates couple to hyperfine eigenstates with $F = 3/2$. In the presence of strong magnetic field $\gtrsim 1\,\text{T}$, as assumed for the existence of GD states, the electronic Zeeman term of Eq.~\eqref{eq:Hspin} dominates the interaction, rendering the matrix in Eq.~\eqref{eq:HspinMat} into near-diagonal form with weak off-diagonal elements. Thus, the eigenstates are close to the basis states $\ket{m_{I_2}|m_{s_1}m_{s_2}}$. In this regime, the eigenstates are energetically split, depending on the total spin orientation $M=m_{s_1}+m_{s_2}$, by several $10\,\text{GHz}$, separating states of parallel and anti-parallel spins. The anti-parallel states are further split by the hyperfine interaction. The parallel spin states can also be expressed as triplet states, while the anti-parallel spin states may be represented by a superposition of singlet and triplet states \cite{PDG_book_CG}, which results in a coupling of these mixed-character spin states to both scattering channels of the interaction potential $V_\text{GD,n}$ in Eq.~\eqref{eq:Vint}.

	\section{Cylindical Giant Dipole States}\label{sec:Meth}
	In order to solve the eigenvalue problem associated with the molecular Hamiltonian in Eq.~\eqref{eq:molHam}, we make use of the Born--Oppenheimer approximation. Separating the total wave function $\Psi(\bvec r, \bvec R) = \chi(\bvec R) \psi(\bvec r; \bvec R)$ yields
	\begin{align}
		\left[H_\text{GD} + H_\text{spin} + V_\text{GD, n}(\bvec r; \bvec R)\right]&\psi_i(\bvec r; \bvec R) =\nonumber\\ =& \epsilon_i(\bvec R) \psi_i(\bvec r; \bvec R)\\
		\left[\frac{\bvec p_\text{n}^2}{2 m_\text{n}} + \varepsilon_i(\bvec R)\right]\chi_k^i(\bvec R) &= E_k^i\chi_k^i(\bvec R)
	\end{align}
	where $\psi(\bvec r; \bvec R)$ describes the electronic wave function that depends parametrically on the nuclear separation $\bvec R$. The corresponding eigenenergies serve as potential energy surfaces (PES) in the equation of the vibrational wave functions $\chi(\bvec R)$. 
	
	To calculate the PES via exact diagonalization, we expand $\psi_i$ into an appropriate basis. In spin space, we choose the eigenbasis of $H_\text{spin}$ discussed in Sec.~\ref{sec:spinSpace}, and in coordinate space we choose the eigenbasis of $H_\text{GD}$. For the chosen field strengths, the frequencies $\omega_x$ and $\omega_y$ of the GD potential, Eq.~\eqref{eq:Vgd}, are nearly equal and justify a polar approximation $\omega_x\simeq\omega_y=\omega_\rho$. Hence, the GD system can be described as a charged isotropic two-dimensional harmonic oscillator in the $(x,y)$-plane in a homogeneous magnetic field along $z$, and a one-dimensional harmonic oscillator in $z$-direction. 
	
	The wave function can thus be written as \cite{Dippel_Giant_Dipole}
	\begin{equation}
		\psi_{n,m,n_z}(\rho, \varphi, z) = R_{nm}(\rho) \frac{\ee^{\ii m \varphi}}{\sqrt{2\pi}} \phi_{n_z}(z) \, \text . \label{eq:GDwave}
	\end{equation}
	The first term is the radial wave function of the two-dimensional isotropic harmonic oscillator
	\begin{gather}
		R_{nm}(\xi) = \sqrt{\frac{2 \mu \Omega_\rho (n + |m|)!}{n!(|m|!)^2}} \nonumber\\ \times
		\ee^{- \frac 1 2 \xi^2}(\xi)^{|m|} {}_1F_1(n; |m| + 1; \xi^2) \,,
	\end{gather}
	with $\xi=\sqrt{\mu \Omega_\rho}\rho$ and $\Omega_\rho^2=\omega_\rho^2+\omega_c^2$ where the magnetic field enters in the modified cyclotron frequency $\omega_c = B (m_p-m_e)/m_pm_e$. The $_1F_1(a; b;z)$ are confluent hypergeometric functions. The second factor in Eq.~\eqref{eq:GDwave} are the eigenfunctions of $\hat L_z$, and last term is the wave function of the one-dimensional harmonic oscillator
	\begin{equation}
		\phi_{n_z}(z) = \left(\frac{\mu \omega_z}{\pi}\right)^{1/4} \frac{1}{\sqrt{2^n \, n!}} H_n\left(\sqrt{\mu \omega_z}z\right) \ee^{- \frac 1 2 \mu \omega_z  z^2}\, \text ,
	\end{equation}
	with the Hermite polynomials $H_n$. The eigenenergies of $H_\text{GD}$ are hence given by
	\begin{equation}
		E_{n,m,n_z} = \Omega_\rho(2 n + |m| + 1) + \frac {\omega_c} 2 m + \omega_z \left(n_z + \frac 1 2\right) \, \text .
	\end{equation}

	Solving the GD system without approximations as in Refs.~\cite{Dippel_Giant_Dipole,Markus_erstes} leads to a decomposition into two independent harmonic oscillators with one large eigenfrequency $\omega_+ =4.13\times 10^{11} \,\text{rad/s}$ and one small eigenfrequency $\omega_-=2.23\times 10^8\,\text{rad/s}$, while the $z$-component has an intermediate spacing of $\omega_z = 1.35 \times 10^9 \, \text{rad/s}$. In contrast, the cylindrical approximation results in three different eigenfrequencies for $n$ and $m$ and depending on the sign of $m$, which we label $\omega_n$ and $\omega_{m\pm}$. The smallest of these frequencies is $\omega_{m-}=2.23\times 10^8\,\text{rad/s}$, which can be identified with $\omega_-$ of the exact solution. The two other frequencies $\omega_n=4.13635\times 10^{11}\,\text{rad/s}$ and $\omega_{m+} = 4.13412 \times 10^{11}\,\text{rad/s}$ are much larger and differ only by one $\omega_{m-}$. Applied to a cylindrical system, the large eigenfrequency $\omega_+$ of the exact solution is degenerate, resulting either from an excitation with $m>0$ or an excitation $n$ combined with $m<0$. Due to the large eigenfrequencies, exceeding the dissociation threshold of the GD potential, excitations of $n$ or positive values of $m$ have been excluded from our exact diagonalization calculations.

	By varying the maximum quantum numbers $n_{z,\text{max}}$ and $m_\text{min}<0$ we reached a relative accuracy of $10^{-4}$ by including angular excitations $m\in[0,-60]$ and $z$-excitations $n_z\in[0,30]$, resulting in a total basis set of $1891$ GD states. Recent studies indicate that this method still implies a numerical error of about $10\%$ \cite{Stielow_2017_Greens}, however, it predicts the correct shape and overall topology of the PESs within reasonable computation times. 
	
	In spin space, the diagonalization can also be simplified due to the large Zeeman splitting in strong magnetic fields. In the matrix representation introduced in Sec.~\ref{sec:spinSpace}, the projection operators of the scattering interaction, Eq.~\eqref{eq:Vint}, read
	\begin{equation}
		\mathbb P_1 = \frac{1}{\sqrt 2}\begin{pmatrix} 0& 0& 0& 0\\ 0& 1& 1& 0\\ 0& -1& 1& 0\\ 0& 0& 0& 0\end{pmatrix}, \  \text{and} \  \mathbb P_3 = \begin{pmatrix} 1& 0& 0& 0\\ 0& 0& 0& 0\\ 0& 0& 0& 0\\ 0& 0& 0& 1\end{pmatrix},
	\end{equation}
	respectively. Therefore, a pure triplet eigenstate of $H_\text{spin}$ does not couple to any other spin state, and both triplet states can be diagonalized independently, providing a route for direct comparison with Ref.~\cite{Markus_erstes}. In contrast, the two mixed-character spin states $\ket{+1/2|\uparrow\downarrow}$ and $\ket{+1/2|\downarrow\uparrow}$ are strongly coupled by $\mathbb P_1$ and have to be considered simultaneously, resulting in $2 \times 1891 = 3782$ basis states necessary for diagonalization.

	\section{Potential energy surfaces}\label{sec:PES}
\begin{figure}[htb]
		\includegraphics[width=\columnwidth]{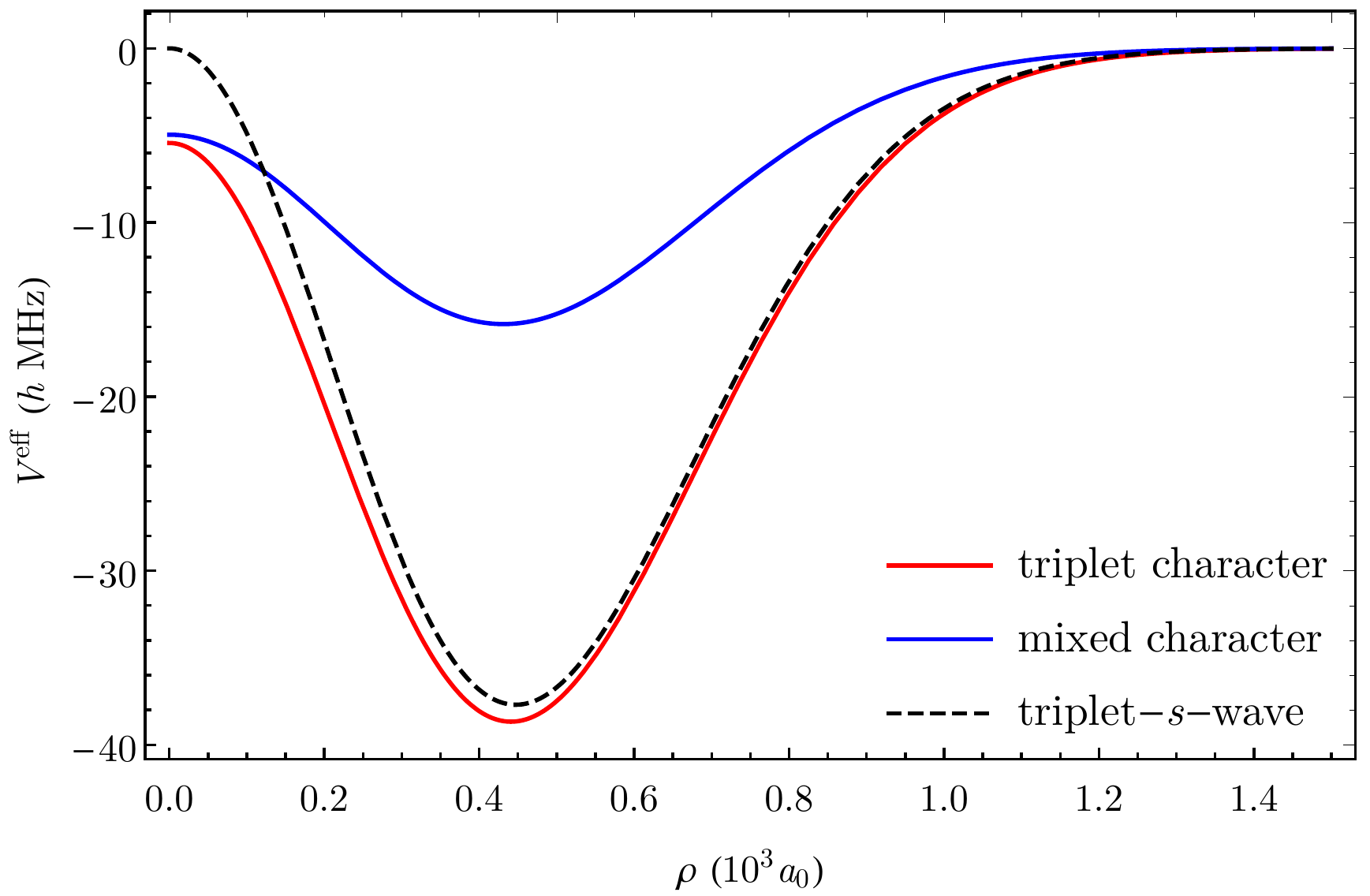}  
		\caption{The PES of the triplet states (red) is widely dominated by $s$-wave scattering. The PES of the mixed-character states (blue) reaches only about $40 \%$ of the triplet curve due to the coupling to the repulsive singlet $s$-wave scattering channel.}\label{fig:pertComp}
	\end{figure}

	A fast and direct approach to study the impact of different interaction terms is to treat them in first-order perturbation theory, where the $s$-wave interaction is proportional to the probability density and the $p$-wave to its derivative of the considered state. Figure~\ref{fig:pertComp} shows the PESs of the $m=-1$ excited GD state in both triplet and mixed-character spin configuration compared to a pure triplet $s$-wave approach as in Ref.~\cite{Markus_erstes}.
	
	For better comparability, all PESs are plotted without the constants offsets by $\varepsilon_\text{GD}$ and $\varepsilon_\text{spin}$. Hence, only one triplet-character and one mixed-character curve are displayed. Far from the origin, the triplet-character curve (red) is dominated by the singlet scattering channel and follows the pure triplet $s$-wave model (dashed). Close to $\rho=0$, $p$-wave contributions become relevant which result in a finite energy plateau. 
	
	The mixed-character states (blue) couple to both the attractive triplet $s$-wave-scattering channel as well as to the repulsive singlet $s$-wave scattering with equal strength, resulting in a shallow PES with only about $40 \%$ depth.  The $p$-wave scattering is approximately equal for both spin states, hence the mixed-character PES converges to nearly the same value as the triplet PES for $\rho \to 0$. Compared to Rydberg molecules, the electron momentum is about one order of magnitude smaller. Therefore, no shape resonances can be observed and the impact of the $p$-wave scattering contributes only a minor correction.

\begin{figure}[htb]
		\includegraphics[width=\columnwidth]{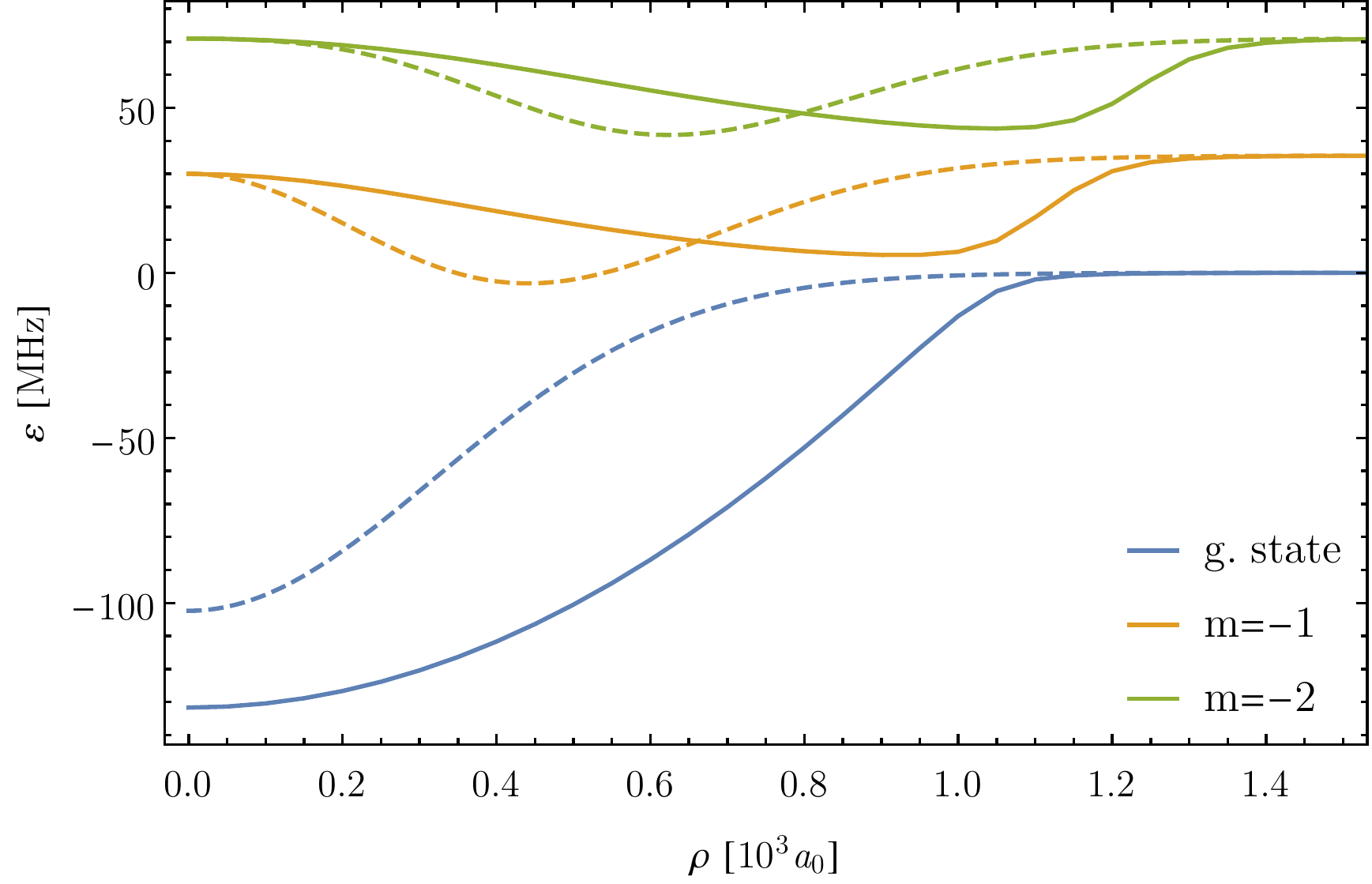}  
		\caption{PESs obtained by exact diagonalization (solid curves) are significantly deformed compared to perturbative results (dashed curves) due to couplings to higher excited states. The GD ground state PES (blue) is broadened to a parabolic shape. The first excited states are dominated by the coupling to $m=-1$ (orange) and $m=-2$ (green) GD states, respectively.}\label{fig:exDiag3}
	\end{figure}

	Similar to Ref.~\cite{Markus_erstes} we observe a notable change in the shape of the PESs when numerically diagonalizing the system. Figure~\ref{fig:exDiag3} shows a comparison of the first three GD states in triplet spin configuration ($\ket{\downarrow\downarrow}$).
	
	The GD ground-state PES is deformed from a Gaussian to a broader parabolic shape, which has also been reported in Ref.~\cite{Markus_erstes}. Varying the included excited states reveals that these shifts originate from a coupling to higher $z$-excited states with the same parity.  We further observe a large shift of the first excited PESs away from the origin, originating from couplings to $m$-excited GD states. The same effect is also observed for higher $m$-excited states.

\begin{figure}[htb]
		\includegraphics[width=\columnwidth]{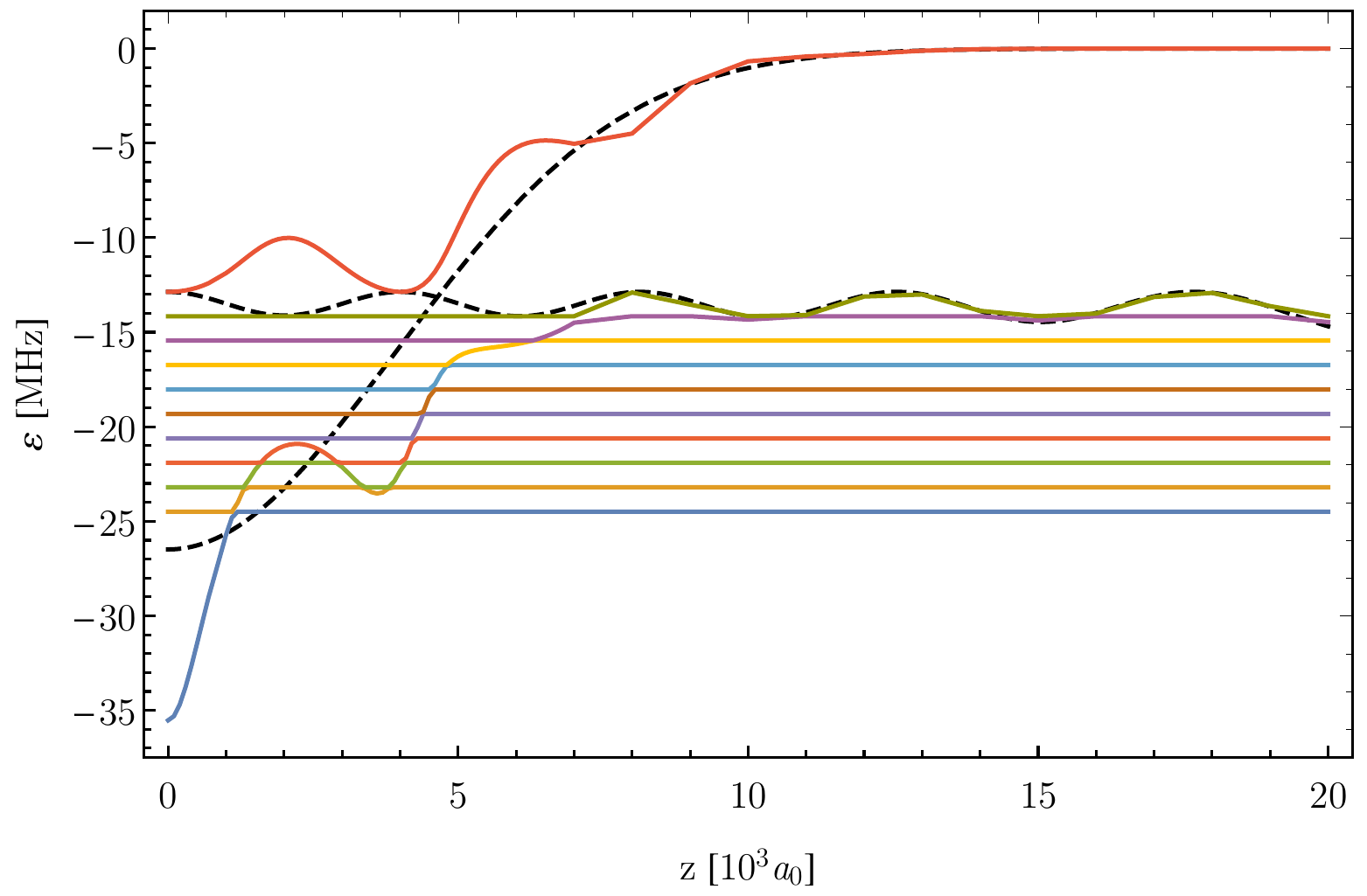}
		\caption{Avoided crossings of the mixed-spin $\ket{+1/2|\uparrow\downarrow}$ GD ground-state PES (outermost blue and red) with GD excited lower spin states $\ket{+1/2|\downarrow\uparrow}$ at $\rho = 300\,a_0$. The dashed curves indicate the PES shapes in case of vanishing coupling, obtained by first-order perturbation theory, of the ground state and $\{m=-2,n_z = 9\}$ state, respectively.}\label{fig:slicedState}
	\end{figure}

	In addition to the avoided crossings of $z$-excited states reported in Ref.~\cite{Markus_erstes} we also observe avoided crossings between different spin states. The two mixed-character spin states are separated energetically only by the hyperfine interaction, which equals roughly 9 GD $z$-excitations or 60 $m$-excitations. Hence, the GD ground-state PES of the spin state $\ket{+ 1 /2|\uparrow \downarrow}$ couples to several excited GD states of the $\ket{+ 1/2|\downarrow \uparrow}$ state, with quantum numbers ranging from $\{m=-56, n_z=0\}$ to $\{m = -2, n_z = 9\}$. Consequently, the original ground-state PES is split into an upper and a lower branch, separated by a manifold of shallow PESs, as shown in Fig.~\ref{fig:slicedState}. As the Born--Oppenheimer approximation breaks down in the vicinity of avoided crossings, we do not expect any of these sliced states to be strongly affected by non-adiabatic transition dynamics.

	\section{Rovibrational states}\label{sec:Rovi}
\begin{figure}[htb]
		\includegraphics[width=\columnwidth]{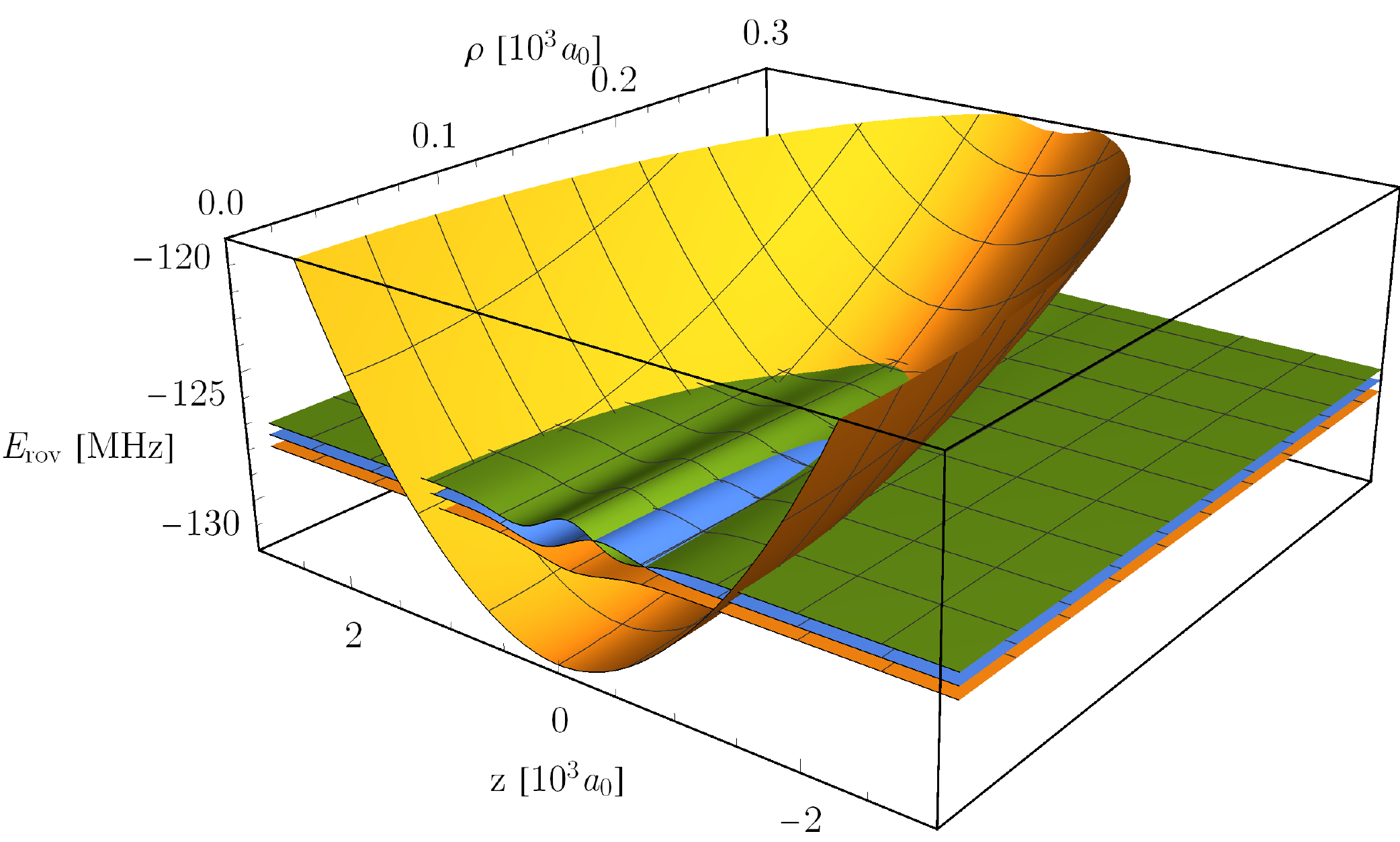}  
		\caption{Rovibrational ground state and first two excited states supported by a triplet ground-state PES. The composite image shows the PES (orange) in the $(R,Z)$-plane. On the vertical axis the wave functions are scaled to arbitrary units with offsets equal to their binding energy inside the PES.}\label{fig:rovi3D}
	\end{figure}

	Based on the Born--Oppenheimer PES described in the previous section, it is possible to calculate the rotational and vibrational spectrum of GD molecules through the Schr\"odinger equation
	\begin{equation}
		\left[\frac{\bvec P_\text{n}^2}{2 m_\text{n}} + \varepsilon_i(\bvec R)\right]\chi_k^i(\bvec R) = E_k^i \, \chi_k^i(\bvec R) \, \text .
	\end{equation}
	The azimuthal symmetry of the PES carries over to the rovibrational wave functions, allowing the ansatz
	\begin{equation}
		\chi(R, \Phi, Z) = \frac{\ee^{\ii \, \nu_\phi \Phi}}{\sqrt{2\pi}} \frac{u(R, Z)}{\sqrt R}
	\end{equation}
	in cylindrical coordinates $\{R, \Phi, Z\}$ of the perturber atom. We solve the corresponding residual Schr\"odinger equation for $u(R, Z)$ using a fourth-order finite-difference method. The wave functions and binding energies of the rovibrational ground state as well as the first two excited states supported by the triplet GD ground-state PES are shown in Fig.~\ref{fig:rovi3D}.
 	The states are deeply bound inside the PES close to the minimum. The ground state is located at a binding energy of $128.2 \, h \, \text{MHz}$ with a spacing of $0.5 \, h \, \text{MHz}$ to the first excited state, meaning the PES may support several hundred excited states. In Ref.~\cite{Markus_erstes} it was shown that in this regime the potential is near harmonic, resulting, again, in a cylindrical harmonic oscillator. Therefore, to all excited states one can assign quantum numbers $\{\nu_R,\nu_\phi,\nu_Z\}$, which are the principal quantum number, the angular quantum number, and the $Z$ quantum number, respectively. These quantum numbers can best be identified by the structure of the electronic probability density in the $(R, Z)$-plane. 
 	
	\begin{figure}[htb]
		\includegraphics[width=\columnwidth]{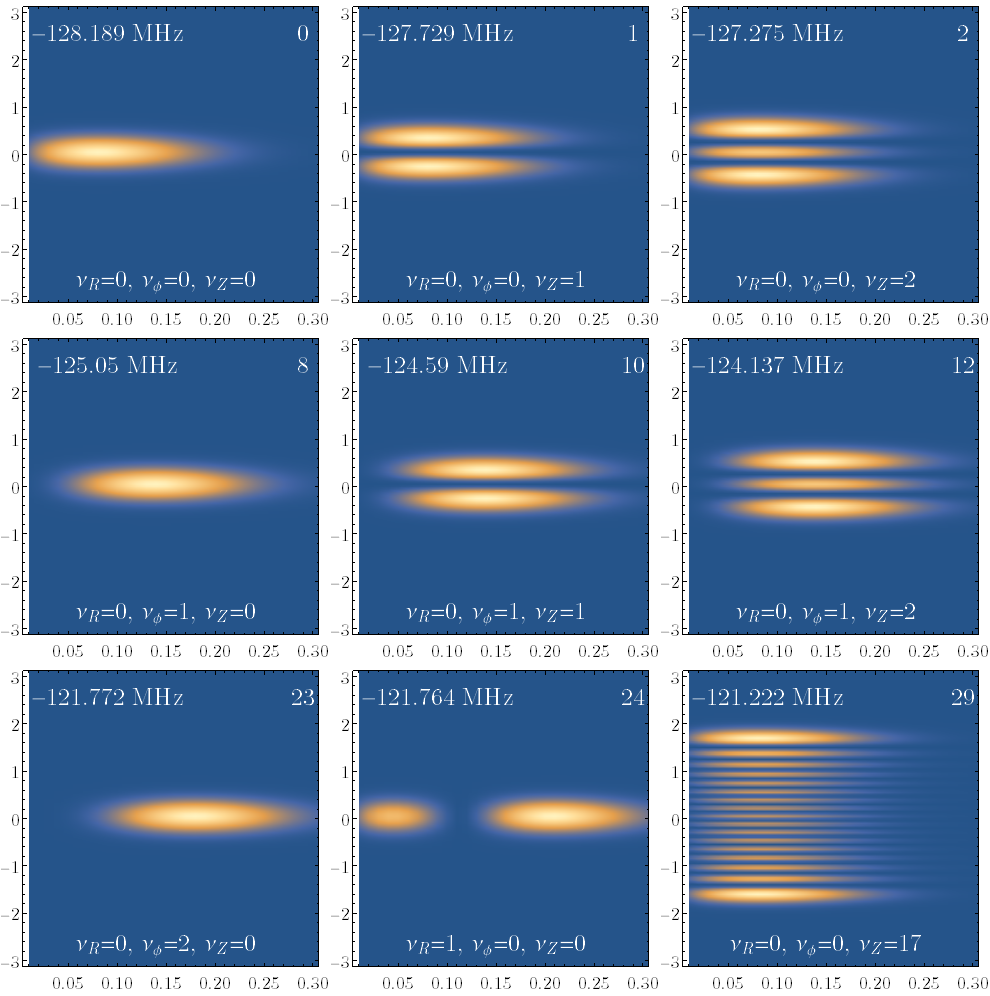}
		\caption{Probability densities of selected rovibrational states of the triplet GD ground state in the $R$-$Z$-plane ($R$ horizonzally, $Z$ vertically, each in units of $10^3 \, a_0$). The numbers in the top right corners indicate the total excitation number in energetic order, while the lower set labels the corresponding harmonic excitation numbers of a three-dimensional cylindrical harmonic oscillator.}\label{fig:densPlots}
	\end{figure}
	
	Probability densities of some characteristic states are displayed in Fig. \ref{fig:densPlots}. The $\nu_Z$ excitations can be identified by the number of nodes in the $Z$-direction. Due to the strong spatial confinement of the GD states in radial direction compared to the $Z$-direction, the spacing in $Z$ is almost one order of magnitude smaller than the angular spacing. Consequently, the first non-$Z$-excited state in energetic order is the 8th excited state with $\nu_\phi = 1$. It can be identified by the characteristic shift away from $R=0$ and the absence of any nodes along $Z$.

 	The next higher angular-excited state is the 23rd state in total, which is even stronger shifted outwards. In a purely harmonic oscillator potential this state would be energetically degenerate with the $\nu_R=1$ state. Our calculations show small variations, indicating the deviations from a purely harmonic model. The harmonic approximation holds for both triplet and mixed GD ground states as well as for the single wells of simple GD excited PESs. An exception is the previously mentioned sliced ground state of the 3rd spin state. The corresponding PES is so strongly deformed and shallow, that its ground state only has a binding energy of $15.9\,h\,\text{MHz}$ with rapidly decreasing spacings on the order of $1\,h\,\text{MHz}$. The excited states are heavily influenced by the finite depth of the PES and, due to the weak binding energies, may dissociate thermally.

	\section{Conclusion}\label{sec:conc}
	We have systematically explored the impact of fine structure and hyperfine couplings in ultra-long-range giant-dipole molecules formed between a highly excited hydrogen atom and a $5^2S_{1/2}$ ground state rubidium atom. Similar to previous works, we have treated the electron-perturber interaction within a Fermi pseudopotential approach that includes both the $s$-wave and $p$-wave singlet and triplet scattering channels. 
	
	The couplings introduced by the newly considered spin interactions strongly influence the PESs obtained within the Born--Oppenheimer approximation. 
	We have found a new type of mixed-spin molecules, which are bound by both the singlet and triplet channel of the Fermi-pseudopotential. 
	Our calculations further indicate interactions between the two possible mixed-spin state configarations, induced by the perturber interaction. 
	The cross couplings result in avoided crossings, highly deforming the topology of one certain spin state.
	In consequence, the molecular interaction may lead to a dissociation of GD atoms prepared in this spin state.
	
	Numerical solutions of the vibrational equations predict deeply bound, stable molecular states supported by both triplet and mixed-spin character PES with binding energies up to hundred MHz. The resulting rovibrational level spacings are in the order of 1--2 MHz. The vibrational spectrum is similar to ordinary Rydberg molecules and may be observed in the laboratory upon the successful preparation of GD states.

	\section{Acknowledgments}
	T. S. acknowledges financial support from ``Evangelisches Studienwerk Villigst''. We further acknowledge support from Deutsche Forschungsgemeinschaft (DFG) within the SPP 1929 ``Giant interactions in Rydberg systems''.

\end{document}